\documentclass[preprint2]{aastex}

\usepackage{graphics,latexsym,psfig}

\def\etal{{et\,al.}}
\def\degs{\ifmmode ^{\circ}\else$^{\circ}$\fi}
\def\amin{\ifmmode ^{\prime}\else$^{\prime}$\fi}
\def\asec{\ifmmode ^{\prime\prime}\else$^{\prime\prime}$\fi}
\def\fss{\hbox{$.\!\!^{\rm s}$}}        
\def\h{$^{\rm h}$}\def\m{$^{\rm m}$}

\newcommand{\xmm}{XMM-{\it Newton}}

\shortauthors{Greiner et al.}
\shorttitle{ROSAT SSS with Chandra}

\begin{document}

\title{Supersoft X-ray sources in M31: II. ROSAT-detected supersoft 
sources in the ROSAT, Chandra and XMM eras}

\author{J. Greiner\altaffilmark{1}}
  \affil{Max-Planck-Institut f\"ur Extraterrestrische Physik, 
                 Giessenbachstra\ss{}e, 85748 Garching, Germany}

\author{R. DiStefano\altaffilmark{2}, A. Kong, F. Primini}
\affil{Harvard-Smithsonian Center for Astrophysics, Cambridge, 
MA 02138, U.S.A.}

\altaffiltext{1}{jcg@mpe.mpg.de}
\altaffiltext{2}{Also: Department of Physics and Astronomy, Tufts University, 
   Medford, MA 02155, U.S.A.}



\begin{abstract}

We have performed Chandra observations during the past 3 years 
of 5 of the M31 supersoft X-ray sources discovered with ROSAT.
Surprisingly, only one of these sources has been detected, despite a 
predicted detection of about 20-80 counts for these sources.
This has motivated a thorough check of the ROSAT M31 survey I data,
including a relaxation of the hardness ratio requirement used
to select supersoft sources.
This increases the number of supersoft sources identified in survey I by 7.
We then carried out a comparison with the ROSAT M31 survey II dataset 
which had
hitherto not been explicitly investigated for supersoft X-ray sources.
We find that most of the ROSAT survey I sources are not detected,
and only two new supersoft sources are identified.
The low detection rate in the ROSAT survey II and our Chandra
observations implies that the variability time scale of
supersoft sources is a few months. If the majority of these
sources are close-binary supersoft sources with shell hydrogen burning,
it further implies that half of 
these sources predominantly experience large mass transfer rates.

\end{abstract}



\section{Introduction}

Observations during the past decade have suggested the definition
a new class of sources. Luminous supersoft X-ray sources (SSSs)  
have 
luminosities in the range $10^{35}-10^{38}$ erg/s
and $kT$ in the range $20-80$ eV, with no hard
X-ray component of comparable luminosity.
Some SSSs are simply hot white dwarfs (e.g., post novae), or 
pre-white-dwarfs (in planetary nebulae). 
What is most
intriguing  about SSSs, however, is the fact that the physical
nature of a majority of the sources with optical IDs
is not yet understood.
These more mysterious sources include the prototypes
CAL 83 and CAL 87, discovered with Einstein (Long, Helfand, \& Grabelsky 1981),
and more numerously with ROSAT (e.g. Greiner 2000).
The most promising explanation for the majority of the sources 
invokes quasi-steady nuclear burning
of matter accreting onto the surface of a white dwarf (WD) to generate
these systems' prodigious flux (see, e.g., van den Heuvel \etal\ 1992). 
There is indirect evidence in favor of these models for several of
the sources.
The binary sources which are so luminous that nuclear-burning models seem to be
required, will be referred to as close-binary supersoft sources (CBSS).

Observing supersoft sources  in M31 has the advantage that several
questions can be attacked more easily as compared to the local
sources (including those in the Magellanic Clouds): (i) What is
the spatial distribution
over the galaxy and possible correlations with different environment?
(ii) What is the size of the population including the ratio of SSS to 
 other types of low-mass X-ray binaries? (iii) What is the variability
pattern (if any) and duty cycle?
All these questions can help in providing clues to the nature of the sources.

   \begin{deluxetable}{rlccccc}
   \tablewidth{16cm}
   \tablecaption{New SSS from the ROSAT PSPC}
\tablehead{
    No &  ~~~~~Name & Coordinate  & Error    &
                   count rate & HR1 & HR2 \\
       &        &   (2000.0)  & (\asec)&
                   (cts/ksec) &
                       &      }
\startdata
  \noalign{\smallskip}
 \multicolumn{7}{c}{First observation during PSPC Survey I} \\
 4& RX J0039.3+4047~       &   00\h39\m21\fss4 +40\degs47\amin41\asec & 42
       &  0.26$\pm$0.23 & $-$0.89$\pm$0.10 & $-$0.29$\pm$0.65  \\
 6& RX J0039.7+4030~       &   00\h39\m47\fss1 +40\degs30\amin05\asec & 15 
       &  2.03$\pm$0.30 & $-$0.85$\pm$0.10 & $-$0.83$\pm$0.53  \\ 
 7& RX J0039.8+4053~       &   00\h39\m50\fss4 +40\degs53\amin38\asec & 23 
       &  1.07$\pm$0.25 & $-$0.75$\pm$0.18 & ~~0.44$\pm$0.97   \\ 
 9& RX J0040.4+4013~       &   00\h40\m28\fss6 +40\degs13\amin44\asec & 23
       &  0.50$\pm$0.27 & $-$0.85$\pm$0.14 & ~~0.72$\pm$1.00  \\ 
14& RX J0042.7+4107~       &   00\h42\m44\fss9 +41\degs07\amin18\asec & 22 
       &  1.04$\pm$0.31 & $-$0.89$\pm$0.16 & $-$0.65$\pm$1.00  \\
17&  RX J0044.2+4117~       &   00\h44\m14\fss0 +41\degs17\amin57\asec & 34 
       &  0.95$\pm$0.35 & $-$0.97$\pm$0.25 & $-$0.58$\pm$0.53  \\
23&  RX J0047.6+4159~       &   00\h47\m42\fss3 +41\degs59\amin59\asec & 36 
       &  1.23$\pm$0.44 & $-$0.82$\pm$0.28 & $-$0.20$\pm$1.00  \\ 
 \noalign{\smallskip}
  \multicolumn{7}{c}{First observation during serendipituous Pointing} \\
21&  RX J0047.4+4157~       &   00\h47\m27\fss2 +41\degs57\amin34\asec & 25
       &  0.60$\pm$0.18 & $-$0.98$\pm$0.21 & ~~0.00$\pm$0.30  \\
\enddata
\vspace{-0.3cm}
           \label{newsssi}
\end{deluxetable}


ROSAT has observed the full disk of the M31 galaxy (about 6.5 deg$^2$) twice.
A ROSAT PSPC mosaic of 6 contiguous pointings with an exposure time of 
25 ksec each was performed in July 1991 (first M31 survey; Supper \etal\ 1997).
A second survey 
was made in July/August 1992 and January 1993 (Supper \etal\ 2001).
Until now, only the first survey has been investigated systematically for SSS
(Greiner \etal\ 1996a).

This paper is the second in a series 
dealing with SSSs in M31, and in particular with
their variability properties. The first
(Di Stefano \etal\ 2004)
concentrated on the analysis of several sets of {\it Chandra}
data: (1) $3$ separate $15$ ksec observations of each of $3$ disk
fields, and (2) a $40$ ksec ACIS observation of the bulge with the
backside-illuminated (BI) chips, combined
with information gleaned from $2$ years of regular ACIS front-side
illuminated (FI) chips monitoring of the bulge.
In fact the disk fields were 
observed such that the locations of 5 
ROSAT SSSs (\#2, 3, 12, 19, 20) were covered by the BI chips
which exhibit enhanced sensitivity for very soft X-rays.
Another four SSS (\#1, 14, 24, 25)
are covered by chance coincidence with the FI 
chips due to the field rotation between the different epochs. 
The results of that paper relevant to this second paper
can be summarized as follows:
\begin{itemize}
\item 
That paper established that only one of the ROSAT-discovered sources,
RX J0038.6+4020, was detected by {\it Chandra}.

\item No new supersoft source obeying the same criteria as
those applied for the selection in ROSAT data, has been found in
any of these Chandra pointings. However, 
with a modified hardness ratio criterion a total of
$16$ new SSSs that are not associated with foreground
or background objects, and which are therefore likely members of M31, 
 were discovered in the disk fields.
Not all of these $16$ were luminous enough to have been detected by ROSAT;
$6$ provided fewer than $20$ counts. Furthermore,
some appear to be hard enough not to have been selected as SSSs
using the procedures applied to the ROSAT M31 survey data.
Nevertheless, at least $3$ of the sources with more than $20$ counts
would have likely been selected as ROSAT SSSs.
Interestingly enough, it could be established that $2$ of these
$3$ sources are transient by comparing the flux at these
positions among different {\it Chandra}
pointings or by studying data taken with {\it XMM-Newton}.

\item The bulge of M31 is rich in high-luminosity SSSs. By comparing
among different {\it Chandra}
pointings or by studying data taken with {\it XMM-Newton}
it has been found that 12 of 16 bulge sources are transient, and one
additional source is highly variable.
\end{itemize}

The non-detection of 4 out of 5 ROSAT-discovered supersoft X-ray sources,
combined with the failure of finding any new sources with similarly
soft X-ray spectra appears to be puzzling for at least two reasons.
First, if one assumes sources with constant brightness, could it be 
possible that the non-detection with Chandra is due to spurious detections
with ROSAT? Second, if one assumes sources with variable X-ray emission,
why do we not detect as many new sources with Chandra as we miss because
they faded away between the ROSAT detection and the Chandra observation?

To answer these questions, we have embarked on a comprehensive
re-analysis of the ROSAT data. In particular, in
this paper we slightly revise the hardness ratio criterion
used to select supersoft sources in the ROSAT 
survey I data (\S 2), we analyze
the second ROSAT PSPC survey for SSS with the same criteria (\S 4) 
plus serendipitous PSPC observations (\S 5), and
we present the first survey of ROSAT supersoft sources
with the {\it Chandra} Observatory (\S 3), and for completeness also
include the  public \xmm\ observations (\S 6). We finally discuss
the variability in (\S 7).

\section{ROSAT PSPC survey I}

The search for supersoft sources in the M31 ROSAT data has been
done so far only on the survey I data. The hardness ratio criterion
HR1 + $\sigma_{\rm HR1} \le -0.80$ (where HR1 is defined as the 
normalized count difference 
(N$_{\rm 50-200}$ -- N$_{\rm 10-40}$)/(N$_{\rm 10-40}$ + N$_{\rm 50-200}$), 
with N$_{\rm a-b}$ 
denoting the number of counts in the PSPC between channels a  and b
(with the approximate conversion of channel/100 $\approx$ energy in keV)
had been applied.  A total of 15 sources were found 
(Greiner \etal\ 1996a, Supper \etal\ 1997). 
This hardness ratio criterion had been copied from a similar search done
for the Magellanic Clouds and the whole PSPC all-sky survey.
For those searches, contamination with cataclysmic binaries in
the Magellanic Clouds or 
local F and G type stars (for the all-sky survey) was a problem
which was mitigated by applying a very strict hardness ratio criterion.
For M31, this problem does not exist, so it is worthwhile to
reconsider the hardness ratio criterion for the survey I data.

How much can we relax the hardness ratio criterion?
Some SSS may be hydrogen-burning white dwarfs, and thus
may reach effective temperatures of
up to 70--80 eV. At a mean galactic foreground absorbing column of 
6$\times$10$^{20}$ cm$^{-2}$ (Dickey \& Lockman 1990) and allowing for
a similar M31 intrinsic absorption, this translates into a hardness
ratio as low as HR1 $\sim$ 0. On the other hand, supernova remnants can have 
hardness ratios as low as HR1 $\sim$ --0.3, so we chose to be
not contaminated by known source types.
Thus, we conservatively adapt HR1 = --0.5 as the new criterion, thus 
ensuing that no other class of sources is included.

The result of relaxing the
hardness ratio criterion to HR1 + $\sigma_{\rm HR1} \le -0.5$
for selecting sources from the M31 ROSAT  survey I
results in (only)  7 additional sources (Tab. \ref{newsssi}) with
respect to the 15 sources obtained with the earlier selection of 
HR1 + $\sigma_{\rm HR1} \le -0.8$ (Greiner \etal\ 1996a, Supper \etal\ 1997).
None of these 7 new sources has a known long-wavelength (optical, 
infrared or radio) counterpart, supporting our claim that these new sources 
have the same nature as the earlier selected 15 sources.
This brings the number of ``canonical'' ROSAT supersoft sources to 22.

Kahabka (1999) has made a different selection to also include
possible supersoft sources which are located behind a substantial absorbing
column. He applied the criteria HR1 $<$ +0.9 and 
HR1 + $\sigma_{\rm HR1} \le -0.1$, and thereby selected 26 additional sources. 
This was motivated by the galactic supersoft source 
RX J0925.7-4758 (Motch \etal\ 1994). However, 8 of the 26 newly selected
objects have been identified with foreground stars or supernova remnants
(Kahabka 1999). Another source is likely a foreground cataclysmic variable.
While Kahabka (1999) argues that the remaining sources are absorbed
supersoft X-ray sources, there is also the possibility that they
are of similar nature as the already identified objects. We therefore
did not include them in the present {\it Chandra} study, but only mention
that we covered 6 objects of his sample with Chandra, two of which are detected
(the bright bulge source RX J0042.8+4115 and RX J0047.6+4132).
For the 4 non-detections no statement about X-ray variability can be made
due to the harder spectra as compared to the canonical supersoft sources, 
and the less favourable ROSAT PSPC to ACIS conversion rate (see below and 
Tab. 2).

\section{Chandra observations in 2000/2001}

Full details of the {\it Chandra} observations of M31 are given elsewhere
(DiStefano \etal\ 2004), so in addition to the results given in the
introduction, we repeat here only the few relevant points.
In order to cover the 5 ROSAT sources (\#2, 3, 12, 19, 20) 
in each of the three different epochs,
we arranged the pointing directions of the S3 chip such that the field 
of view rotated around the center
of the S3 chip, and not the aim point (see Fig. 1). As the field of view
rotated from one epoch to the next, it also covered four other 
supersoft sources  (\#1, 14, 24, 25)
with one of the front-side illuminated chips (marked
with $g$ in Tab. 2).
Given that the {\it Chandra} S3 chip is a factor of two more sensitive than 
the ROSAT PSPC for supersoft sources (at $kT \sim 40$ eV and the low 
foreground absorption towards M31), each of the 15 ksec observations 
was expected to provide of the order of 20--80 counts from 
each supersoft source.


Surprisingly, only one of the {\it ROSAT} sources 
(\#3)
was detected during the {\it Chandra} observations (DiStefano \etal\ 2004). 
Upper limits for the other sources 
were derived at the 2$\sigma$ confidence level, using the full ACIS-S  
energy range (0.25--7 keV), since the background is anyway dominated 
by the soft end of the spectrum. 
The count rates for the detections and the upper limits are summarized
in Tab. 2.

\begin{figure}
\vbox{\psfig{figure=m31_1_epoch1+2+3_ROS_s.ps,width=8.1cm,%
         bbllx=3.08cm,bblly=10.5cm,bburx=18.7cm,bbury=26.2cm,clip=}}\par
\caption{Optical image of the southern part of M31 
  covering one of the three Chandra fields with the location
  of the six ACIS detectors overlaid for each of the three epochs
  (green = 1. epoch, blue = 2. epoch, red = 3. epoch).
  Note that the rotation of the field of view was arranged to happen
  around the center of the S3 chip, and not around the aim point.
  This leads to different off-axis angles of a given source during
  different epochs and explains why our upper limits in Tab. 2
  are usually worse than the on-axis sensitivity of about 6$\times$10$^{-4}$
  cts/s.
  Open white circles are the detected sources with the circle radius
  being proportional to the detected count rate. The three red filled
  dots denote the locations of three ROSAT-discovered supersoft X-ray sources
  (\#1, 2, 3).
  The two of those located within the S3 chip (\#2, 3)
  were covered in each of
  the three epochs, whereas the third (\#1)
  was only covered in the first.
  Only one of these ROSAT supersoft sources (\#3) 
  has been detected with ACIS
  (open circle overlapping with one red dot), and this one only during 
  the first two epochs. The pattern for the two other fields is similar.}
\end{figure}

\section{ROSAT survey II}

\subsection{The data}

The strong X-ray variability implied by the {\it Chandra} results motivated us
to investigate the 22 canonical ROSAT supersoft sources in the 
second ROSAT PSPC survey. This second ROSAT survey was performed in
July/August 1992, January 1993 and July/August 1993, 
and consisted of 96
different pointings with 2.5 ksec each, offset from each other
by about 10 arcmin 
between each other. After merging all these 96 individual pointings,
the second PSPC survey provides a much higher spatial homogeneity
as compared to the 6 survey-I pointings with 25 ksec each, 
and hence
a higher sensitivity in the outer regions of the M31 disk. In addition,
less area of M31 is lost in the second survey due to occultation
by the PSPC window support structure which is an important effect
for the first survey. While the limiting sensitivity in the 0.1--2.0 keV
range is $5\times 10^{-15}$ erg cm$^{-2}$ s$^{-1}$ in the first
survey and $7\times 10^{-15}$ erg cm$^{-2}$ s$^{-1}$ in the second
survey, it is important to keep in mind the above differences which lead
to a substantially different spatial sensitivity pattern across the M31 disk
between the two surveys.

\begin{figure*}
\vbox{\psfig{figure=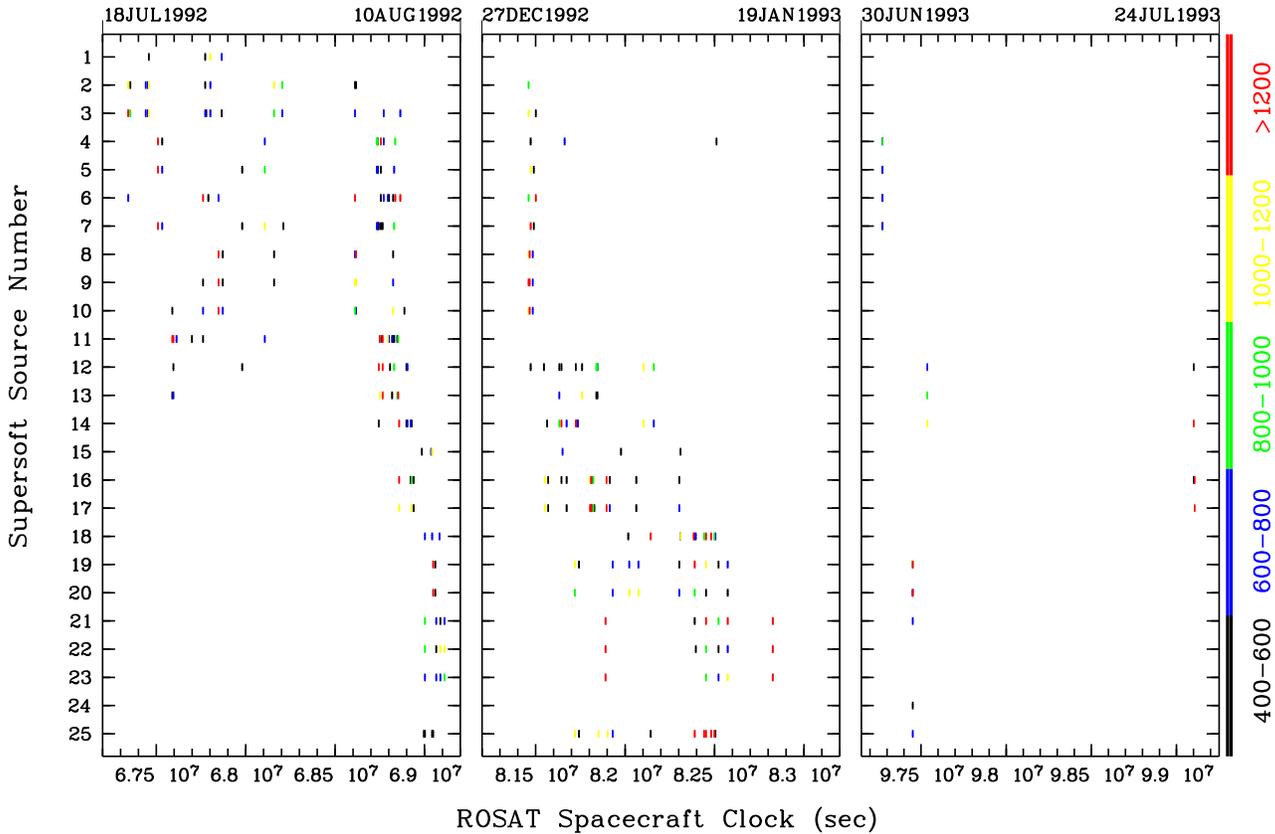,width=17.cm,angle=270,%
         bbllx=2.2cm,bblly=1.cm,bburx=19.6cm,bbury=27.2cm,clip=}}\par
\caption{Temporal sequence of the individual observation intervals 
   of M31 during the second ROSAT survey. This survey has been
   conducted in two main observation epochs, namely the south-eastern 
   part of M31
   during July/August 1992 and the north-western part during December 1992
   to January 1993. For a few pointings, the exposures were completed
   only in June/July 1993, marking a third observation epoch.
   Except for three sources (\#1, 11, 24), the exposure spreads over
   more than one exposure epoch. Shown as color-coded dashes is the
   effective exposure time at the sky location of the 25 supersoft sources
   for all the 96 individual pointed observations. The effective exposure
   has been computed by applying two factors to the on-axis, nominal 
   exposure time:
   (1) the vignetting correction, i.e. the decrease of the effective area with
    off-axis angle, and 
   (2) the square of the ratio of the radius of the point spread function (PSF)
    at the given off-axis angle to that on-axis (for 0.4 keV and 90\%
    encircled energy) which is a correction for the decreasing source 
    detection probability at larger off-axis angles due to the larger
    background area covered by the PSF. Effective exposure times below
    400 sec have been suppressed. The detection of a supersoft source
    with a brightness similar to that seen in the first ROSAT survey
    requires a minimum effective exposure of 4000 sec for the brightest 
    sources (6, 15, 16, 18)
    and $\sim$40\,000 sec for the faintest (1, 4, 5, 11). }
\end{figure*}

A comparison of the source tables of the two surveys
(Supper \etal\ 1997, 2001) shows that only 
 three (\#3,6,18) of the original 15+7=22
 ROSAT SSS have also been detected in the second ROSAT survey.
 Furthermore, only one new SSS (\#24) has been discovered in the
 second ROSAT survey. In order to investigate this in more detail,
 we have used the merged data set of 
Supper \etal\ (2001) and re-investigated the locations of the SSS from
the first survey by searching
 the map and maximum-likelihood detection maps
for SSS at fainter levels than the 4$\sigma$ list of
Supper \etal\ (2001). We re-discover one source (\#15) at the 3$\sigma$ 
level which had fallen below the 4$\sigma$ threshold of the second survey.
We also detect the White \etal\ (1995) transient (\#25) which had not been 
seen in the first survey.
With the one new SSS detection (\#24) and including the 
White \etal\ (1995) transient, 
this results in two new source detections in survey II, and
the sample of SSS in M31 increases to 24 sources (Tab. 2).

Finally, we derived upper limits for 
those sources which have not been detected. Upper limits have been
determined by fitting a Gaussian profile with the width corresponding to the
mean width of the point spread function of the merged pointings
to the known positions, taking into account the vignetting and
effective exposure time, and are given at the 2$\sigma$ level in Tab. 2.

\subsection{Understanding the large fraction of non-detections}

At first glance, this may cause doubts on the quality of the
data and/or analysis. However, we have been very careful in checking
these causes, and are convinced that these causes can be excluded.
First, the original data analysis leading to the merged intensity and
exposure maps of both surveys have been done by the same person
with the same software within less than one year apart (R. Supper in
1996/1997). Second, the majority of the hard X-ray sources are
re-discovered, so if it were a technical problem, then the soft response
would have to have suffered. Given the subsequent non-detections with 
{\it Chandra} one would cast more doubts on the first survey than the second. 
However, there have been many observations of other soft sources
after 1991, including non-interacting white dwarfs and ``monitoring''
observations of soft sources for calibration purposes which show that
the soft response remained very stable until the end of the PSPC life.

The most likely effect leading to the non-detection of the
survey I sources is the ``stretched'' time sampling of the second survey 
in conjunction with intrinsic X-ray variability of the SSS.
This survey was primarily done in three 2-month long 
exposure epochs (in the following called EP1, EP2 and EP3),
separated by 6 months each (between July 1992 and August 1993).
Typically, each of the 96 observations of 2.5 ksec is spread over 2 days.
However, for 13 out of the 96 pointings in the second survey the exposure
was split over two of these three epochs 
(either EP1-EP2 or EP2-EP3), and in two cases even over
three epochs. 
The important fact to realize is that even when an observation was done
within 2 days, it would not be sufficient to detect a SSS. Instead, at least
2 such observations are required for the brightest sources (\#6,15,16,18),
and up to 15 observations for the faintest sources (\#1,4,5,11).
Fig. 2 shows the actual sampling for all SSS and demonstrates that
for most of these sources the survey II exposure is spread over at least
3 weeks. In fact, only 3 sources are observed within one exposure period
(\#1 and \#11 in EP1 = July/August 1992 and \#24 in EP3 = July/August 1993),
while 9 sources are observed over two exposure periods and 12 even over all 
three exposure periods. Only if a SSS was constant over 6-8 months,
i.e. over two exposure periods (either EP1-EP2 or EP2-EP3) or even
12 months (all three EPs) it had a chance to be detected during the
second ROSAT survey.
If, on the contrary, the variability time scale of SSS is shorter than 
6 months, but longer than 3 weeks, only a fraction of the total survey II 
exposure would contribute to the potential detectability. From the 
detailed source coverage by the 96 individual pointings (Fig. 2)
we determine that 
(i) 16 sources received enough exposure within a 3-week
interval to be detectable at their survey-I count rate, out of
which 6 have indeed been detected;
(ii) 3 sources (\#1,4,5) are not detected due to too short exposure
if they remained constant, and
(iii) 6 sources (\#2, 8, 9, 10, 11, 21) required the full survey exposure,
  out of which 1 source was detected.

\section{Serendipitous ROSAT observations}

There have been three long (more than 15 ksec) PSPC observations
of M31 in the same time frame as Survey I and II. While one of these
(Observation-ID 600245)
does not cover any of our SSS, the other two observations do cover 
3 and 9 SSS, respectively. The former observation (Obs-ID 600244) was performed
between January 2--30, 1993 for a total of 35.86 ksec, the other 
(Obs-ID 600121)
between January 5 and February 5, 1992 for a total of 44.73 ksec.
The latter observation is particularly interesting because 
it happened before the second PSPC survey.

A source detection (within the EXSAS package; Zimmermann \etal\ 1994)
was applied, including first a mask creation to screen all the parts of 
the image where the
support structure of the PSPC entrance window affects the
detectability of X-ray photons, secondly a map detection
("sliding window") to find and remove all sources in order to thirdly
produce a background map with a bi-cubic spline fit to the
resulting image. Finally, a maximum likelihood algorithm is applied
to the data (e.g.,  Cruddace, Hasinger \& Schmitt 1988) in three separate
PHA channel ranges. 
For the sources which are not detected, 2$\sigma$ upper limits (Tab. 2)
are computed in the 0.1--0.4 keV range as described above.

None of the three sources covered by the January 1993 observation 
is detected. However, due to the large off-axis angles of these
sources the upper limits are all above the brightness of these sources
during the survey I. That is, these upper limits are consistent with
no variability.

For the other observation (ID 600121 in 1992), which covered 9 SSS, 
three are detected; in all cases at a level similar to
the survey I intensity. Since all these three sources are detected in
the 1992 observation, i.e. about 6 months after the first and
before the second PSPC survey, it re-enforces the earlier interpretation
that the SSS as found in the first PSPC survey are all real.
The upper limits for another 4 sources are again high enough to be
consistent with no variability, 
and two sources have faded (\#20,23).

Applying the revised hardness ratio criterion to these three
pointed observations, plus ignoring sources with bright (up to $V$ = 18 mag)
stars within their error box to avoid bright foreground stars 
(Greiner \etal\ 1996a) we find 1 new supersoft source
(\#21 in Tabs. 1 and 2). This brings the total sample of ROSAT 
SSS in M31 to 25.

\section{\xmm\ observations}

The bulge and disk of M31 were observed by \xmm\ several times  
between 2000 and 2002. In particular, the central 15\amin\ area was observed 4
times (2000 June, 2000 December, 2001 June and 2002 January; see Shirey
\etal\ 2001; Osborne \etal\ 2001; Trudolyubov \etal\ 2002a), while 4
fields covering the 
northern and southern regions of the galaxy were visited by \xmm\ 
in 2002 January (for instance, see Trudolyubov \etal\ 2002b).
All data were taken with the three detectors (pn, MOS-1 and MOS-2) of the
European Photon Imaging Camera (EPIC). The exposure time for the disk
fields is about 60 ksec each while for the central region, the exposure time
varies from 13 ksec to 60 ksec. The archival event
lists were reprocessed and filtered with the {\it XMM-Newton} Science
Analysis Software (SAS v5.4.1). We examined background flares of
each observation and rejected intervals with high background level.
Only data in 0.2--12 keV were used for the analysis.

The (2$\sigma$) upper limits (Tab. 2) have been determined from the
EPIC PN data by using the  XMMSAS {\em emldetect}
algorithm with an external source list and a maximum likelihood threshold
of zero, thus providing upper limit counts derived from a fit of the
three-dimensional point spread function to the photon distribution.

No unbiased serach for supersoft sources has been performed on the
XMM data.

\section{X-ray Variability}

\subsection{The Results}

Looking at Tab. 2, one can summarize the X-ray variability
of the ROSAT-discovered supersoft sources in M31 as follows:

\begin{itemize}

\item Combining the two ROSAT surveys, we find that
out of the 22 SSS detected during the ROSAT survey I, 18 sources
are not detected during survey II.
Two new SSS (\#24, 25) are discovered relative to survey I.  From the 4 
sources (\#4,6,15,18) detected in both surveys, three 
remained constant while one was rising by a factor of 2.
For 8 of the sources (\#1,4,5,7,9,11,12,17) the upper limits during 
the full survey II are consistent with the
measured count rates during survey I. Thus, about half of the supersoft
sources (the above 8 plus 3 sources which are seen in both surveys at
similar count rate) are (or could be) constant.
In total 10 sources (\#2,8,10,13,14,16,19,20,22,23) have faded by a 
factor of 2--5 on a time scale of one year.

\item Three sources (\#18,19,22) have been detected in the serendipitous
PSPC observation in Jan./Feb. 1992 at intensities very similar to those
measured 6 months earlier during the first PSPC survey. While one of
these sources (\#18) increased in intensity thereafter,
the other two (\#19,22) faded by a 
factor of 3--4 until the exposures of the second PSPC survey (6--12 month
later).

\item The serendipitous PSPC observations provide upper limits for 
two sources (\#20,23) demonstrating that they 
faded by a factor of 3 within 6 months.

\item Including the {\it Chandra} and \xmm\ observations, and thus the
longer time scale of 9--12 years, two of the constant sources (\#3,6) showed
fading by a factor of 5--10, and two of the rising sources (\#24,25) faded by
a factor of 5--25.

\item One of the ROSAT-discovered SSS covered by Chandra observations (\#3) was
``on'' in the first and second set of 15 ksec {\it Chandra} observations, 
but ``off'' in the third. Moreover, the count rate declined by nearly a 
factor of two between the two {\it Chandra} epochs, and the decline
between the ROSAT survey II and the first {\it Chandra} epoch was a factor
of three. This points to variability time scales of (shorter than) three 
months and a short duty-cycle. In fact, this source could be 
similar to the fading source RX J0527.8--6954 (Greiner \etal\ 1996b).

\item The {\it Chandra} observations do not reveal any new supersoft source
 with a hardness ratio and count rate comparable to the ROSAT-discovered 
 sources ($>$20 cts in 15 ksec), but {\it Chandra}'s spatial coverage 
was only 5\% of the M31 disk.

\end{itemize}

In conclusion, when sorted for variability time scale and considering
only variability with an amplitude larger than a factor of 2, we have
\begin{itemize}
\item one source (\#3) which varied over a timescale of 3 months,
\item 7   sources (\#18,19,20,21,22,23,24)  which varied over a timescale of 
   6 months,
\item 7 sources (\#2,8,10,13,14,16,25)  which varied over a timescale of 1 yr,
\item 2 sources (\#6,12)  which varied over a timescale of $>$5 yrs, and
\item 8 sources (\#1,4,5,7,9,11,15,17) for which no statement about 
variability can be made.
\end{itemize}

\subsection{Possible origin of the X-ray variability}

If the majority of these sources are close-binary supersoft sources,
one possible explanation for this rapid variability could be
 photospheric expansion and contraction of the white dwarf envelope,
which can shift the radiation
out of and then back into the X-ray regime. This is the
mechanism suspected to be responsible for the X-ray variability
in RX J0513.9-6951 (Reinsch \etal\ 1996) and
CAL 83 (Greiner \& Di\,Stefano 2002). The interesting point, though,
is that if this were true, 
about half  of these sources (just considering sources with a
variability time scale shorter than 1 year) would operate
at rather high mass transfer rates, corresponding to the
upper limit of the stable H burning regime. One then may ask where
to find the sources with mass transfer rates within the stable burning
region? Whether this is an observational bias (since we preferentially
detect the high-temperature, high-luminosity sources in M31 due to
sensitivity reasons) or can be accomodated in population synthesis
models remains to be evaluated in more detail.

However, we do not know whether all the SSS in M31 are close-binary supersoft
sources. There are several
other alternatives, which also would explain variability:
(1) post-nova SSSs should (and have been observed to) dim over time.
The number of SSSs  which can be post-novae is constrained by
independent estimates of the nova rate;
(2) pre-white dwarfs can re-ignite (the ``born-again"
phenomenon); this happens over time scales short enough that
the associated planetary nebulae should still be visible.
(3) any SSBs that are neutron stars -- they sometimes exhibit low-hard states.
One possible example, though not conclusively identified as a neutron star, is 
1E 1339.8+2837, which switches between high/soft and low/hard states 
(Dotani \etal\ 1999);
(4) variable absorption due e.g. variable mass loss, as the soft X-ray
 emission is very vulnerable to column densities above a few times 
10$^{20}$ cm$^{-2}$.

\subsection{The fraction of novae and recurrent novae}

It is interesting to note that one of the faders (RX J0044.0+4118 = \#16) 
has been optically identified as a classical nova which erupted in 1990
(Nedialkov \etal\ 2002). Thus, one could speculate whether the above
difference in the numbers of faders and risers is due to a fraction
of classical novae. However, observing at a given time (i.e. 1991 or
2000) should show a similar number of novae being on in their soft
X-ray state, unless the supersoft phase of novae is so short and/or rare,
that catching one nova during the ROSAT survey I was a unique
chance coincidence. Indeed, a survey of the X-ray emission of
local and nearby novae has shown that only 3 out of 108 novae have
revealed a supersoft phase (Orio \etal\ 2001), While two more supersoft novae
have been identified in the meantime, the majority have  rather short
supersoft phases, of the order of weeks to few months. 
This line of reasoning would then imply that on statistical
grounds RX J0044.0+4118 is most likely the only
nova in the sample of the ROSAT-discovered M31 supersoft sources.
Thus, we do not think that classical nova
can change the ratio of faders to risers, or that they comprise
a substantial fraction of the ROSAT-discovered M31 supersoft sources.

A similar result is obtained when considering the total nova rate of
$\sim$37 nova per year per M31 disk (Shafter \& Irby 2001). 
Since the ROSAT survey I was done in about 1 month, and the duration
of the supersoft phase in novae is of a similar short time scale
(e.g. Greiner \etal\ 2003), at maximum two of the ROSAT-discovered
M31 SSS should be novae, even if all novae would undergo a supersoft
phase.

The outburst rate of recurrent novae in M31 has been estimated to be only
10\% of the rate of classical novae (Della Valle \& Livio 1996).
While this may be an underestimate due to the lower luminosity 
of recurrent novae and the possible lack of sensitivity for part
of the population, it is clear that recurrent novae cannot explain
the frequency of SSS variability in M31.

\section{Conclusions}

The evidence from ROSAT, Chandra, and XMM is that SSSs tend to
be highly variable, perhaps more variable than any other class of
X-ray binary, most notably the hard sources
comprising  a substantial number of X-ray binaries (Trudolyubov \etal\ 2002b).
A large fraction (30\%) 
of SSS are transients with turn-off or turn-on times on the
order of a few months.
The majority of the sources that have
fallen below detectability limits have not been detected again.
This may argue that the duty cycle is low, while activity
times are on the order of months or years. With an on-time
duration of months, and a duty cycle of, e.g.,
40\%, we have only a ten percent chance of detecting a
source which was "on" during one observation, in a 
second uncorrelated observation, and a 16\% chance of detecting any given
SSS in 2 unrelated observations, consistent with our finding.

 In addition, the spatial coverage of M31 
with Chandra was small (less than 5\%), since the coverage for SSS 
by Chandra is primarily given by the S3 chip.
Thus, the likelihood of detecting new SSS with Chandra was small. 
Population studies
have estimated the total SSS population in M31 to be $\sim 1000$
(Di\thinspace Stefano \& Rappaport 1994). With an assumed duty cycle of
10\% this would correspond to a density of active SSS of 3$\times$10$^{-3}$
arcmin$^{-2}$, or 0.02 per S3 chip.

While it is unlikely that novae are responsible for the strong
X-ray variability in SSS, its physical cause  remains to be 
explained. Both, better sampling of the light curve as well as 
optical identifications and subsequent optical monitoring seem
to be required to deduce insight into the variability mechanism(s).

We finally note that the more frequent Chandra and XMM observations  
over the last three years have revealed a number 
of supersoft  X-ray transients (e.g. Shirey 2001, 
Trudolyubov \etal\ 2002c). While their nature remains to be 
established as well, they support the notion of the strong variability of SSS. 



\acknowledgements JG is particularly grateful to Rodrigo Supper 
for providing the
 merged ROSAT PSPC data of the two M31 surveys, which only allowed to derive
the upper limits presented here.
RDS acknowledges support by NASA/Chandra grant GO1-2022X, 
LTSA grant NAG5-10705 and the Croucher Foundation.


\newpage

\begin{deluxetable}{rlccccccc}
\tablewidth{20cm}
\tablecaption{Count rates or upper limits (in cts/ksec) of ROSAT-detected 
supersoft sources 
as seen with Chandra and \xmm. These count rates are on-axis rates as 
detected in each of the instruments (after correction for effective area), 
so are not normalized. Conversion factors are given
in the notes $b,c,g$. 
Sources \#1-20 and \#22/23 are survey I discoveries, sources \#24/25
are survey II discoveries, and source \#21 was first found in a serendipituous
observation. \vspace*{-0.2cm}}
\tablehead{
\colhead{$\!\!$No$\!\!$} & ~~~Source Name      &
\multicolumn{3}{c}{{ROSAT/PSPC \tablenotemark{a}}} & 
\multicolumn{3}{c}{{Chandra/ACIS \tablenotemark{b}}} & XMM/PN\tablenotemark{c}  \\
 & & \colhead{Survey I} & \colhead{Survey II} &  \colhead{Seren-} &
             \colhead{Epoch I} & \colhead{Epoch II} & \colhead{Epoch III} &\\
 & & Jul 1991 & 1992/1993 & dipitous & 1--5 Nov 2000 & 6--8 Mar 2001 & 3 Jul 2001 & }
\startdata
 1& RX J0037.4+4015~ & 0.31$\pm$0.31 & $<0.40$ & $<$1.15  & $<$0.99\tablenotemark{g} & -- & -- & -- \\
 2& RX J0038.5+4014~ & 0.80$\pm$0.28 & $<0.13$ & $<$2.84 & $<$0.41 & $<$0.35 & $<$0.54 & --\\
 3& RX J0038.6+4020~ & 1.73$\pm$0.29 & 1.69$\pm$0.35&$<$2.66 & 0.95$\pm$0.35&0.58$\pm$0.14 & $<$0.29 & --\\
 4& RX J0039.3+4047~ & 0.26$\pm$0.23 & $<0.26$ & --  & -- & -- & -- &-- \\
 5& RX J0039.6+4054~ & 0.44$\pm$0.44 & $<$0.49 &--& -- & -- & -- & --\\
 6& RX J0039.7+4030~ & 2.03$\pm$0.30 & 1.89$\pm$0.34&-- & -- & -- & -- &$<$1.28$^{\rm t}$ \\
 7& RX J0039.8+4053~ & 1.07$\pm$0.25 & $<$1.08 & --  & -- & -- & -- & --\\
 8& RX J0040.4+4009~ & 0.85$\pm$0.32 & $<$0.20 & --  & -- & -- & -- & --\\
 9& RX J0040.4+4013~ & 0.50$\pm$0.27 & $<$0.23 & --  & -- & -- & -- & --\\
10& RX J0040.7+4015~ & 1.26$\pm$0.32 & $<$0.42 & --  & -- & -- & -- & --\\
11& RX J0041.5+4040~ & 0.32$\pm$0.18 & $<$0.23 &--& -- & -- & -- & --\\
12& RX J0041.8+4059~ & 0.49$\pm$0.24 & $<$0.47 &--& $<$0.14 & $<$0.31 & $<$0.15 & $<$2.01$^{\rm t}$ \\
13& RX J0042.4+4044~ & 1.69$\pm$0.32 & $<$0.32 &--& -- & -- & -- & --\\
14& RX J0042.7+4107~ & 1.04$\pm$0.31 & $<$0.20 & --  & $<$3.47\tablenotemark{g} & -- & $<$3.49\tablenotemark{g} & $<$0.75$^{\rm m}$ \\
15& RX J0043.5+4207~ & 2.15$\pm$0.55 & 2.20$\pm$0.77\tablenotemark{d}&$<$2.07 & -- & -- & -- & --\\
16& RX J0044.0+4118\tablenotemark{e} & 2.46$\pm$0.42 & $<$0.77&$<$2.98 & -- & -- & -- & --\\
17& RX J0044.2+4117~ & 0.95$\pm$0.35 & $<$0.75  & $<$2.68 & -- & -- & -- &-- \\
18& RX J0045.5+4206~ & 3.14$\pm$0.34 & 7.41$\pm$0.66& 3.96$\pm$0.39& -- & -- & -- & $<$7.59$^{\rm m}$ \\
19& RX J0046.2+4144~ & 2.15$\pm$0.39 & $<$0.82& 1.96$\pm$0.34 & $<$1.22 & $<$0.79 & $<$0.14 &-- \\
20& RX J0046.2+4138~ & 1.12$\pm$0.40 & $<$0.34&$<$0.47 & $<$0.22 & $<$0.22 & $<$0.29 &-- \\
21& RX J0047.4+4157~ &  $<$0.17 & 0.38$\pm$0.15 & 0.60$\pm$0.18  & -- & -- & -- & --\\
22& RX J0047.6+4159~ & 1.23$\pm$0.44 & $<$0.54 & 1.28$\pm$0.25  & -- & -- & -- & --\\
23& RX J0047.6+4205~ & 1.05$\pm$0.36 & $<$0.17 & $<$0.32  & -- & -- & -- & --\\
24& RX J0047.8+4135~ &   $<0.81$ & 2.35$\pm$0.80&$<$0.76 & -- & $<$8.10\tablenotemark{g} & -- & --\\
 \noalign{\smallskip}
 \hline
 \noalign{\smallskip}
25&  RX J0045.4+4154\tablenotemark{h} & $<0.22$ &$\!\!$2.77$\pm$0.35\tablenotemark{f}  &--   & $<$3.90\tablenotemark{g} & -- & $<$2.98\tablenotemark{g} & $<$0.51$^{\rm m}$ \\ 
\enddata
\vspace{-0.2cm}
\tablenotetext{a}{
 Upper limits are 2$\sigma$ confidence level in the 0.1--0.4 keV band.
 The serendipitous pointings were done during Jan 2--30, 1993 for the sources
  in the top 3 lines, and Jan 5 -- Feb 5, 1992 for the sources in the lower
  part of the table.}
\tablenotetext{b}{Upper limits are 2$\sigma$ confidence level in the 
  0.25--7 keV band.
  The conversion factor between ROSAT PSPC count rate
  and Chandra ACIS-S count rate (cts/ksec) for these supersoft spectra
  ($kT \sim 40$ eV, $N_{\rm H}$ = 6$\times$10$^{20}$ cm$^{-2}$) is 1:2,
  i.e. the ACIS-S count rate is twice the ROSAT PSPC rate.
  Sources not covered by the corresponding observation are marked with
  a horizontal dash.}
\tablenotetext{c}{The observation dates are different for each source:
   RX J0041.8+4059: 12/13 Jan 2002; RX J0045.4+4154 and RX J0045.5+4206:
   26/27 Jan 2002; RX J0039.7+4030: 24/25 Jan 2002; RX J0042.7+4107: 
  25 Jun 2000, 28 Dec 2000 (upper limit is $<$2.98$^{\rm m}$ cts/ksec), 
  29 Jun 2001 (upper limit is $<$1.09$^{\rm m}$ cts/ksec). The superscripts at 
  the upper limits denote
  the optical blocking filter used: t=thin, m=medium. The conversion factor 
  between ROSAT PSPC count rate and the XMM/EPIC PN count rate for
  supersoft sources (same parameters as above) are 1:6.7 for the thin filter
  and 1:5 for the medium filter.}
\tablenotetext{d}{Detection at the 3$\sigma$ level; this source is not
  marked in Ref. 1 as being detected in both surveys due to the 
  4$\sigma$ detection threshold.}
\tablenotetext{e}{Identified as a classical nova, which erupted in Sep. 1990;
  see Ref. 2.}
\tablenotetext{f}{New estimate, which differs from that given in Ref. 1.}
\tablenotetext{g}{These upper limits are for the front-side illuminated
  CCD chips, for which the ROSAT PSPC to Chandra ACIS-I conversion 
  factor is 1:0.2 only.}
\tablenotetext{h}{This source 
  has been first reported by White \etal\ (1995)
  based on ROSAT HRI observations, and seems to be a recurrent transient, 
  as it also was detected in two Chandra HRC snapshot observations; Ref. 3.}
\tablerefs{
(1) Supper \etal\ (2001),
(2) Nedialkov \etal\ (2002).
(3) Williams \etal\ 2004}
\end{deluxetable}



\begin{thebibliography}{}

\bibitem[]{chs88} 
Cruddace, R.G., Hasinger, G.R. \& Schmitt, J.H.M.M., 1988,
in {\it Astronomy from Large Databases}, eds. F. Murtagh F. \& A. Heck, 
(Garching, ESO publications) p. 177

\bibitem[]{dvl96} Della Valle M., Livio M., 1996, ApJ 473, 240

\bibitem[]{dsr94} Di\,Stefano R., Rappaport S., 1994, 
 ApJ 437, 733

\bibitem[]{dkg03} Di\,Stefano R., Kong A.H., Greiner J., \etal\ 2004,
  ApJ (in press) (Paper I)

\bibitem[]{dag99} Dotani T., Asai K., Greiner J., 1999, PASJ 51, 519


\bibitem[]{gsm96} Greiner J., Supper R., Magnier E.A., 1996a, in Supersoft
 X-ray Sources, ed. J. Greiner, Lecture Notes in Phys. 472, Springer, p. 75

\bibitem[]{gsh96} Greiner J., Schwarz R., Hasinger G., Orio M., 1996b, 
A\&A 312, 88

\bibitem[]{gre00} Greiner J., 2000, New Astr. 5, 137

\bibitem[]{gds02} Greiner J., DiStefano R., 2002, A\&A 387, 944

\bibitem[]{ptk99} Kahabka P., 1999, A\&A 344, 459

\bibitem[]{lhg81} Long K.S., Helfand D.J., Grabelsky D.A., 1981, ApJ 248, 925

\bibitem[]{mhp94} Motch C., Hasinger G., Pietsch W., 1994, A\&A 284, 827

\bibitem[]{nob02} Nedialkov P.L., Orio M., Birkle K., Conselic C., 
 Della Valle M., Greiner J., Magnier E., Tikhonov N.A., 2002, A\&A 389, 439

\bibitem[]{oco01} Orio M., Covington J., \"Ogelman H., 2001, A\&A 373, 542

\bibitem[]{obt01} Osborne J.P., Borozdin K.N., Trudolyubov S.P., \etal\ 2001,
  A\&A 378, 800

\bibitem[]{rtba96} Reinsch K., van Teeseling A., Beuermann K., Abbott T.M.C., 
     1996, A\&A 309, L11

\bibitem[]{rtk00} Reinsch K., van Teeseling A., King A.R., Beuermann K.,  
    2000, A\&A 354, L37

\bibitem[]{shi01} Shafter A.W., Irby B.K., 2001, ApJ  563, 749

\bibitem[]{shy01} Shirey R., 2001, IAU Circ. 7659

\bibitem[]{ssb01} Shirey R., Soria R., Borozdin K.N., \etal\ 2001, 
  A\&A 365, L195

\bibitem[]{shp97} Supper R., Hasinger G., Pietsch W., Tr\"umper J., Jain A.,
         Magnier E.A., Lewin W.H.G., van Paradijs J., 1997, A\&A 317, 328

\bibitem[]{shl01} Supper R., Hasinger G., Lewin W.H.G., Magnier E.A.,
van Paradijs J., Pietsch W., Read A.M., Tr\"umper J.,  2001, A\&A 373, 63

\bibitem[Trudolyubov et al.(2002)]{tbp02a} Trudolyubov S.P.,
Borozdin K.N., Priedhorsky W.C., \etal\
 2002a, ApJ 581, L27

\bibitem[]{tbp02b} Trudolyubov  S.P., Borozdin K.N., Priedhorsky W.C.,
  \etal\ 2002b, ApJ 571, L17

\bibitem[]{tbp02c} Trudolyubov  S.P., Priedhorsky W.C., Borozdin K.,
  \etal\ 2002c, IAU Circ. 7798

\bibitem[]{vdh92} van den Heuvel E.P.J., Bhattacharya D., Nomoto K.,
              Rappaport S.A., 1992, A\&A 262, 97

\bibitem[]{wgh95} White N.E., Giommi P., Heise J., Angelini L., Fantasia S., 
  1995, ApJ 445, L125 

\bibitem[]{wgk03} Williams B.F., Garcia M.R., Kong A.K.H. \etal\ 2004, 
  ApJ (subm., astro-ph/0306421)

\bibitem[]{zbb94}
Zimmermann H.U., Becker W., Belloni T.,  \etal, 1994, MPE report 257

\end{thebibliography}
\end{document}